\input{aipcheck}

\documentclass[
    ,final            
  ]
  {aipproc}

\layoutstyle{6x9}

\begin{document}

\title{Gamma-Ray Burst Early Afterglows}

\classification{95.30.-k, 95.55.-n, 95.85.-e, 97.60.-s}
\keywords      {Gamma-ray burst; afterglow; X-ray; UV; optical; {\em Swift}}

\author{Bing Zhang}{
  address={Department of Physics, University of Nevada Las Vegas, Las
Vegas, NV 89154}
}



\begin{abstract}
The successful launch and operation of NASA's {\em Swift}
Gamma-Ray Burst Explorer open a new era for the multi-wavelength study
of the very early afterglow phase of gamma-ray bursts (GRBs). GRB
early afterglow information is essential to explore the 
unknown physical composition of GRB
jets, the link between the prompt $\gamma$-ray emission and the
afterglow emission, the GRB central engine activity, as well
as the immediate GRB environment. Here I review some of the recent
theoretical efforts to address these problems and describe how the
latest {\em Swift} data give answers to these outstanding questions.
%
%
\end{abstract}

\maketitle


\section{Introduction}

Our understanding of cosmological gamma-ray bursts (GRBs) have been
greatly advanced during the past several years. Before the launch of
the {\em Swift} satellite (on Nov. 20, 2004), there have been a few
outstanding questions in the study of GRBs that call for more definite
answers (see e.g. \cite{ZM04} for more detailed discussions). For
example, where is the prompt gamma-ray emission emitted, at the
external shock just like the afterglows or at an ``inner'' radius due
to shock or magnetic dissipation?  What is the physical composition of
the GRB jets, baryonic or magnetic? If they are baryonic, are there
free neutrons in the fireball?  How does the GRB central engine work?
Does it become dormant when the prompt gamma-ray emission is over?
What is the immediate environment of GRBs, a constant density (ISM)
medium or a massive stellar wind?  Are short duration GRBs different
from the long duration GRBs?  The {\em Swift} satellite
\cite{gehrels}, thanks to its capability of promptly slewing the
on-board X-Ray Telescope (XRT) \cite{burrows} and the UV-Optical
Telescope (UVOT) \cite{roming} to the GRB target triggered by the
Burst Alert Telescope (BAT) \cite{barthelmy}, is an ideal mission to
address these questions. The X-ray afterglow
data\footnote{Theoretically speaking an afterglow is the signal
emitted when the fireball is decelerated by the ambient medium which
lasts much longer than the prompt emission itself. In this sense, some
X-ray signals detected after the prompt gamma-rays (e.g. X-ray flares
\cite{Burrows05}) are not afterglows. Here we follow the convention of
defining an afterglow as the electromagnetic signals detected after
the prompt gamma-ray emission is over.}  retrieved as early as less
than $\sim 100$ s after the GRB trigger would give valuable
information on the transition between the prompt gamma-ray emission
phase and the afterglow phase. In the optical band, it is highly
expected that the early afterglow lightcurve should include the
contribution of a short-lived reverse shock component. The late
afterglows, on the other hand, originate from the forward shock and
therefore reflect the emission from the circumburst medium. The
reverse shock component is therefore very precious since it directly
carries the information of the GRB outflow itself, and is valuable to
diagnose the physical composition of the fireball jets.  By carefully
diagnosing the early afterglow data, one can also retrieve valuable
information of the GRB central engine and the properties of the
circumburst environment. {\em Swift} has indeed been very fruitful in
addressing these problems during the first several months of its
operation. Here I summarize some of the latest theoretical work on GRB
early afterglows, and discuss how the models are compared against the
abundant {\em Swift} data.

\section{Optical band: expectations}

\subsection{The standard forward-reverse shock model}

A generic GRB model \cite{RM92,MR93,MR97} suggests that regardless of
the mechanism of the explosion and the property of the central engine,
a relativistic ejecta (fireball) expanding into the space would be
eventually decelerated by the circumburst medium. Usually a pair of
shocks form, i.e. a long-lived forward shock propagating into the
ambient medium and a short-lived reverse shock propagating into the
fireball itself \cite{MR97,SP99,MR99}. Since the shocked fireball
ejecta is typically 
denser than the shocked medium by a factor of $\sim \Gamma_0$, the
initial Lorentz factor of the fireball, the 
typical emission frequency of the reverse shock component is $\sim
\Gamma_0^2$ times smaller than that of the forward shock
component\footnote{This is valid for the
so-called thin-shell regime, i.e. the burst duration $T$ is shorter
than the fireball deceleration time defined by the total energy $E$
and the density $n$.}. As a result, the reverse shock emission peaks
in the UV/optical/IR band according to the standard theory.

Before {\em Swift} is launched, there have been already extensive
observational efforts using ground-based robotic telescopes to catch
the very early optical afterglow of GRBs. Although most of these
searches only place upper limits, some observations yielded
well-monitored early afterglow lightcurves. In particular, GRB 990123
\cite{Akerlof99} and GRB 021211 \cite{fox03a,li03} show impressively
similar early lightcurve behavior marked by a clear transition from
roughly $\propto t^{-2}$ to roughly $\propto t^{-1}$. GRB 021004 shows
a different behavior \cite{fox03b}. In any case, models invoking the
reverse shock have been proposed to interpret these early optical
behaviors \cite{SP99,MR99,kz03a,wei03}. 

\begin{figure}
  \includegraphics[height=.3\textheight]{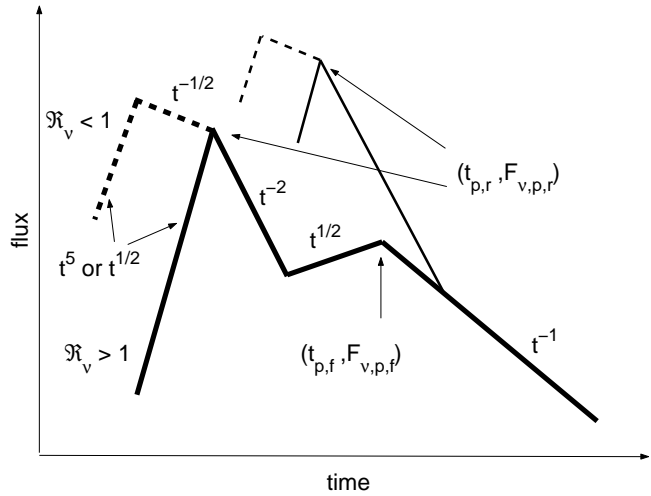} \\
  \caption{Typical early optical afterglow lightcurves (ISM case). Two
types are identified: the re-brightening case (thick), and the
flattening case (thin) which usually requires a strongly magnetized
reverse shock. From \cite{ZKM03}.} 
\end{figure}

A unified treatment of both the reverse shock and the forward shock
components \cite{ZKM03} suggests that if the ambient medium density is
constant (e.g. ISM) in general one expects two lightcurve peaks
(Fig.1). The reverse shock flux peaks at the time when the reverse
shock crosses the shell \cite{K00}, while the forward shock peaks at
the time when the typical synchrotron frequency $\nu_m$ crosses the
optical band \cite{SPN98}. If the shock parameters ($\epsilon_e$,
$\epsilon_B$ and $p$) are similar to each other in both shocks,
generally one expects a lightcurve with two distinct peaks detectable,
and the lightcurve is categorized as the ``re-brightening'' type
\cite{ZKM03} (thick line in Fig.1). On the other hand, since the
ejecta composition could be rather different from that of the
circumburst medium, it is quite 
plausible that the shock parameters are different in both shocks. In
particular, if the central engine is strongly magnetized, the magnetic
fields in the reverse shock region could be (much) stronger than those
in the forward shock region. By introducing a parameter ${\cal R}_B
\equiv B_r/B_f$ (where $B_r$ and $B_f$ are the magnetic field
strengths in the reverse and the forward shock, respectively), it is
found that \cite{ZKM03} the forward shock peak is usually buried
beneath the reverse shock component only when ${\cal R}_B \gg 1$ (see
also \cite{fan02,KP03,MKP04,PK04}). The lightcurve is categorized as
the steep ($\propto t^{-2}$) to flat ($\propto t^{-1}$) transition, or
the ``flattening'' type \cite{ZKM03} (thin line in Fig.1). The
well-studied cases of GRB 990123 and GRB 021211 belong to such a
category, i.e. the reverse shock interpretation requires a strongly
magnetized central engine. The ``re-brightening'' type lightcurve has
been used to interpret GRB 021004 \cite{kz03a} and GRB 041219A
\cite{FZW05a}. 

The early optical afterglow lightcurves in a wind medium has been
studied in \cite{CL00,wu03,kz03b,kmz04,zwd05}. It is noticed that in
such an environment the reverse shock region usually overlaps the
prompt gamma-ray photon beam \cite{FZW05b}. The Inverse Compton
cooling is therefore significant \cite{B05}, which would modify the
reverse shock emission behavior significantly and leads to additional
observational signatures in the GeV range \cite{FZW05b}.

\subsection{Diagosing GRB fireball composition}

{\it Baryonic or magnetic?}
Since a strongly magnetized central engine was inferred at least in
GRB 990123 and GRB 021211 \cite{ZKM03,fan02,KP03}, a direct question
is whether the GRB outflows are dominated by strong magnetic fields
(i.e. a Poynting flux). In other words, how large is the $\sigma$
parameter, which is defined as the ratio between the Poynting flux and
baryonic kinetic energy flux in the outflow? Previous treatments of
the reverse shock dynamics/radiation are purely
hydrodynamical. Magnetic fields are included only through an
equipartion parameter $\epsilon_B$. In order to treat magnetic fields
self-consistently and study the reverse shock emission for an outflow
with a wide range of $\sigma$ values, one needs to start with the MHD
shock jump conditions. This has been done in \cite{ZK05} (see also
\cite{FWW04} for the discussions for $\sigma \leq 1$). According to
\cite{ZK05}, the reverse shock peak flux increases with $\sigma$ when
$\sigma \leq 1$. This is mainly because for $\sigma \leq 1$ the
dynamics of the flow is essentially not modified compared
with the purely hydrodynamical case. On the other hand, $\epsilon_B$
in the reverse shock region keeps increaseing with $\sigma$, so that
the synchrotron emission becomes progressively stronger. It reaches a
peak around $\sigma \sim 1$, with an ${\cal R}_B \sim (3
\epsilon_{B,f})^{-1/2} \sim 18 \epsilon_{B,f,-3}$. This is roughly the
${\cal R}_B$ value inferred in GRB 990123 \cite{ZKM03}, which explains why
the bright optical flash seen in GRB 990123 is rare since it requires
the most optimized $\sigma$ value (around unity) to achieve a large
${\cal R}_B$. As $\sigma$ becomes larger than unity, the flow becomes
Poynting flux dominated. The reverse shock becomes progressively
weaker and it disappears when $\sigma$ is as high as $\sim 100$. The
reason is that given a same initial Lorentz factor $\Gamma_0$, the
typical internal energy density in the forward shock is at most
$e_{\rm 2,max}= 4 \Gamma_0^2 m_p c^2$, and the pressure at the contact
discontinuity is at most $p_{\rm 2,max}=e_{\rm 2,max}/3$. As $\sigma$
becomes higher and higher, the magnetic pressure behind the contact
discontinuity ($p_3$) would become larger and larger, and eventually
exceeds $p_{\rm 2,max}$ so that no reverse shock could
form. Notice that the disappearance of reverse shock in the
high-$\sigma$ regime is not due to that a shock can not exist in a
high-$\sigma$ flow (in fact, the shock suppression factor essentially
does not decrease in the high-$\sigma$ regime, \cite{ZK05}), but is
rather due to the maximum available forward shock pressure defined by
$\Gamma_0$. A large fraction of {\em Swift} bursts were not seen by
UVOT at very early epochs \cite{Roming05}. This suggests that the
reverse shock emission is strongly suppressed. Among other
possibilities, a Poynting-flux-dominated model is a plausible
interpretation. Within this scenario, the early afterglow is expected
to be intrinsically dim initially since the bulk of the energy
contained in the magnetic fields is not transferred to the ISM by the
time the reverse shock disappears \cite{ZK05}. This seems to be
consistent with the fact that some early-UVOT-dark bursts have
a relatively faint X-ray afterglow flux level at 1 hr after the
trigger, resulting in a very high apparent gamma-ray emission
efficiency \cite{Roming05}. 

{\it Neutron-rich fireball?}
If the fireball is dominated by baryons, it has been suggested that a
substantial fraction of baryons are free neutrons and the decay of
these neutrons would lead to interesting observing features in the
early afterglow phase \cite{derishev99,B03}. Detailed calculations
have been carried out \cite{FZW05c}. For an ISM-type medium, the
leading neutrons decay, drag the ISM and shock into the
medium. This ``neutron-decay-trail ejecta'' is decelerated by
the ISM, and is eventually caught up with by the trailing proton
shell \cite{FZW05c}. As a result, the neutron signature in the ISM
case is essentially an energy injection signature, which has been
extensively modeled before (e.g. \cite{ZM02}). In a wind medium, the
direct emission from the neutron decay trail may become
important. However, the ``overlapping effect'' with the gamma-ray
photon flow makes the signature not very significant \cite{FZW05c}.

Besides leaving imprints in the early afterglows, free neutrons 
would also play an important role in the internal shock phase
(e.g. \cite{FW04,rossi04}). The neutron-rich internal shocks could
result in strong optical emission associated with the prompt
gamma-rays due to much weaker synchrotron self-absorption at the
neutron decay radius, which could interpret the apparent association 
between gamma-ray and optical emission detected in GRB 041219A
(\cite{FZW05a,vestrand05,blake05}).

{\it Swift data.} Because of the previous positive detections of the
optical flashes in 
GRB 990123 and GRB 021211 \cite{Akerlof99,fox03a,li03} and intensive
theoretical investigations of the GRB reverse shock, it has been
highly expected that the {\em Swift} UVOT would record many nice early
afterglow lightcurves that allow us to study the reverse shock in
greater detail. Indeed a fraction of {\em Swift}-triggered bursts has
early optical afterglows recorded by UVOT and other groundbased robotic
telescopes, and the reverse shock component has been identified in GRB
041219A \cite{FZW05a} and possibly also in GRB 050525A
\cite{blustin05,SD05}. Yet it is still out of one's expectation that
the majority of the UVOT bursts are ``dark'' from the very beginning,
despite of the prompt slews and the deep exposures \cite{Roming05}. 
It might be that at least some GRBs are Poynting flux dominated, so
that the reverse shock emission component is suppressed \cite{ZK05}.



\section{X-ray band: surprises}

\begin{figure}
  \includegraphics[height=.45\textheight,angle=-90]{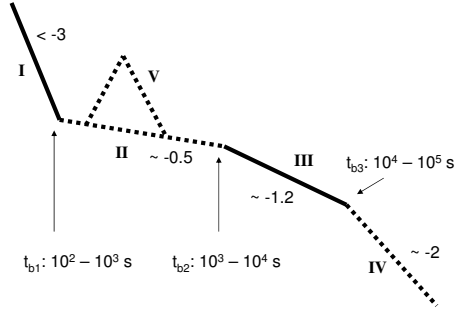}
  \caption{A canonical early X-ray afterglow lightcurve inferred from the {\em
Swift} XRT observations. From \cite{Zhang05}.}
\end{figure}

Contrary to the optical band, the X-ray band was only sparsely studied
in the early afterglow phase before the launch of {\em Swift}. This is
because only little early X-ray data were available, and because the
reverse shock emission is not expected to be important in the band.
It turns out that {\em Swift} XRT detected an X-ray afterglow for
essentially every burst (in contrast to UVOT), and it brings several
surprises to the GRB workers, since the new phenomena are not
straightforwardly expected in the pre-{\em Swift} era.

After inspecting a sample of early X-ray afterglow data
(e.g. \cite{Chincarini05,Nousek05}, one can draw a synthetic X-ray
afterglow lightcurve \cite{Zhang05} (see Fig.2). This lightcurve
includes 5 components: I. an early steep decay; II. a follow-up
shallower-than-normal decay; III. a normal decay; IV. a jet break; and
V. one or more X-ray flares. Three time breaks ($t_{b1}$, $t_{b2}$ and
$t_{b3}$) separate the segments I, II, III and IV. Not every
burst has all these components, but they are all common features. The
segments III (normal decay) and IV (jet break) are expected based on
previous late optical afterglow observations. However, the other three
components are regarded as ``surprises''.

{\it Surprise 1: steep decay.} A rapid decay component is evident in
the early X-ray lightcurves of the majority of the {\em Swift} bursts
\cite{Tagliaferri05}. In some bursts this component is smoothly
connected to the spectrally-extrapolated BAT lightcurve
\cite{Barthelmy05-1}, and it could be generally interpreted as the
``tail emission'' of the prompt emission due to the so-called
``curvature effect'' \cite{KP00}. In some cases, a mismatch between
the steep decay and the BAT lightcurve is evident, but 
it could be well the tail emission of the X-ray flares
\cite{Burrows05}. The curvature-effect suggests that the temporal
decay index $\alpha$ and the spectral index $\beta$ should satisfy
$\alpha = 2 + \beta$ \cite{KP00}. In order to accomodate the data,
additional effects (e.g. shifting the time zero point to the last
pulse in the prompt emission or the X-ray flare, substracting the
underlying forward shock component, etc.) are needed \cite{Zhang05}. 
Alternatively, the steep decay may be also a result of the central
engine activity \cite{FW05,Zhang05}.
This distinct component suggests that the GRB prompt emission (and the
X-ray flare emission) originate from a different location than that of
the afterglow, i.e. likely at
an ``internal'' radius within the deceleration radius \cite{RM94}.
The large contrast between the prompt emission component and the
afterglow component (connected with the steep decay component) also
suggests a very high gamma-ray emission efficiency \cite{Zhang05}.

{\it Surprise 2: shallow-than-normal decay.} In a good fraction of
the {\em Swift} bursts, the decay slope in the afterglow phase (after
the steep tail emission ends) is shallower than what is expected in
the standard afterglow model with a constant energy. The total energy
in the fireball needs to increase with time, so that the fireball is
continuously refreshed for a much longer time than the burst
duration \cite{Zhang05}. This corresponds to Segment II in Fig.2. 
There are three physical possibilities that could give rise to such a
refreshed shock. 1. The central engine keeps pumping energy with a
reduced rate, e.g. $L(t) \propto t^{-q}$ \cite{ZM01}; 2. The energy
injection from the central engine is brief but the ejecta have a wide
range of Lorentz factors with a power-law distribution \cite{RM98};
3. The outflow is Poynting-flux-dominated, so that the magnetic field
takes a longer time to be transferred into the medium \cite{ZK05}.
A successful model must interpret why the injection index $q \sim 0.5$
is inferred from the data \cite{Zhang05,Nousek05} and why there is a
well-defined epoch when the injection ceases abruptly (see
\cite{Zhang05} for more discussions).

{\it Surprise 3: X-ray flares.} Yet another surprise is the 
X-ray flares detected in nearly half of the {\em Swift} GRBs
\cite{Burrows05,Falcone05,Romano05}. Although a weak flare (e.g. in
GRB 050406) may be interpreted as the synchrotron self Compton emission
in the reverse shock region \cite{Kobayashi05} under well balanced
conditions, the general properties of the flares (e.g. the large
amplitude in GRB 050502B, rapid rising and falling lightcurves, 
more than one flares in one burst) strongly suggest that the correct
mechanism is the late central engine activity \cite{Zhang05} (see also
\cite{Burrows05,FW05}). Even more surprisingly, after the breakthrough
of localizing short, hard GRBs and building a close link between the
short bursts and the compact star merger models
\cite{Gehrels05,Barthelmy05-2}, extensive X-ray flares are discovered
in the short burst GRB 050724 \cite{Barthelmy05-2}. The observed late
central engine activity in compact star mergers pose great challenge
to the merger modelers. In particular, it is argued that the central
engine mechanism to power flares in the merger scenario must be of
magnetic origin, and the X-ray flares are expected to be linearly
polarized \cite{FZP05}. Numerical simulations with full MHD effects
(e.g. \cite{Proga03}) are desirable in the models of the GRB central
engine. 

\section{Speculations}

With the current abundant early afterglow data collected by {\em
Swift} and other ground-based optical telescopes, it is now evident
that the GRB early afterglow phase is more complicated than what one
could imagine in the pre-{\em Swift} era. The simplest reverse$+$ 
forward shock picture, although applicable in the interpretations, is
inconclusive.

One speculation is that we seem to be collecting evidence that at
least some GRBs are strongly magnetized or even
Poynting-flux-dominated. The tight early UVOT upper limits
\cite{Roming05}, the apparent high gamma-ray efficiency 
in some bursts \cite{Roming05}, the flat injection phase identified in
the X-ray afterglow lightcurves of a group of bursts
\cite{Zhang05,Nousek05}, as well as the argument regarding the X-ray
flare mechanism \cite{FZP05}, all seem to be consistent with such a
picture (e.g. \cite{ZK05}). More data and more detailed modeling are
needed to verify such a speculation.

Another speculation is related to X-ray flares. Before the discovery
of the X-ray flares following GRBs, there has been no serious thought about
such softer and weaker flares at later times. With what we observe now,
one may boldly imagine the existence of even softer flares. Are there
optical flares associated with the X-ray flares or even not associated
with the X-ray flares? In particular, is the bright 9-mag optical
flash detected in GRB 990123 actually an optical flare due to the GRB
central engine activity\cite{Wu05}? This comes back to the original
suggestion of 
the internal shock origin of the optical flash \cite{MR99}. With the
new information collected recently, such an intriguing possibility
is worth re-investigating.


\begin{theacknowledgments}
I thank fruitful collaborations with S. Kobayashi, Y. Z. Fan, 
P. M\'esz\'aros, J. Dyks, D. M. Wei and D. Proga on the theoretical
topics covered in this review. I also gratefully acknowledge all the
colleagues in the {\em Swift} team with whom I had numerous inspiring 
discussions. In particular, I'd like to thank N. Gehrels, D. Burrows,
J. Nousek, and P. Roming for extensive communications and
encouragements. I also like to thank the organizers of the Torun
meeting who made the conference successful. This work is supported by
NASA NNG04GD51G and a NASA Swift GI (Cycle 1) Program.
\end{theacknowledgments}



\bibliographystyle{aipproc}   

\bibliography{sample}

\begin{thebibliography}{9}

\bibitem{ZM04}
B. Zhang, and P. M\'esz\'aros, \emph{Int. J. Mod. Phys. A},
\textbf{19}, 2385-2472 (2004)

\bibitem{gehrels}
N. Gehrels et al., \emph{Astrophy. J.}, \textbf{611}, 1005-1020
(2004) 

\bibitem{burrows}
D. N. Burrows et al., \emph{Proc. SPIE}, \textbf{5165}, 201-216 (2004)

\bibitem{roming}
P. Roming et al., \emph{Proc. SPIE}, \textbf{5165}, 262-276 (2004)

\bibitem{barthelmy}
S. D. Barthelmy et al., \emph{Proc. SPIE}, \textbf{5165}, 175-189 (2004)

\bibitem{Burrows05}
D. N. Burrows et al., \emph{Science}, in press, astro-ph/0506130 (2005)

\bibitem{RM92}
M. J. Rees and P. M\'esz\'aros, \emph{Mon. Not. R. Astron. Soc.},
\textbf{258}, 41P-43P (1992) 

\bibitem{MR93}
P. M\'esz\'aros and M. J. Rees, \emph{Astrophys. J.}, \textbf{405},
278-284 (1993) 

\bibitem{MR97}
P. M\'esz\'aros and M. J. Rees, \emph{Astrophys. J.}, \textbf{476},
232-237 (1997)

\bibitem{SP99}
R. Sari and T. Piran, \emph{Astrophys. J.}, \textbf{517}, L109-L112
(1999) 

\bibitem{MR99}
P. M\'esz\'aros and M. J. Rees, \emph{Mon. Not. R. Astron. Soc.},
\textbf{306}, L39-L43 (1999) 

\bibitem{Akerlof99}
C. Akerlof et al. \emph{Nature}, \textbf{398}, 400-402 (1999)

\bibitem{fox03a}
D. Fox et al. \emph{Astrophys. J.}, \textbf{586}, L5-L8 (2003)

\bibitem{li03}
W. Li et al. \emph{Astrophys. J.}, \textbf{586}, L9-L12 (2003)

\bibitem{fox03b}
D. Fox et al. \emph{Nature}, \textbf{422}, 284-286 (2003)

\bibitem{kz03a}
S. Kobayashi and B. Zhang, \emph{Astrophys. J.}, \textbf{582}, L75-L78
(2003) 

\bibitem{wei03}
D. M. Wei, \emph{Astron. Astrophys.}, \textbf{402}, L9-L12 (2003)

\bibitem{ZKM03}
B. Zhang, S. Kobayashi and P. M\'esz\'aros, \emph{Astrophys. J.},
\textbf{595}, 950-954 (2003)

\bibitem{K00}
S. Kobayashi, \emph{Astrophys. J.}, \textbf{545}, 807-812 (2000)

\bibitem{SPN98}
R. Sari, T. Piran and R. Narayan, \emph{Astrophys. J.}, \textbf{497},
L17-L20 (1998) 

\bibitem{fan02}
Y. Z. Fan, Z. G. Dai, Y. F. Huang and T. Lu, \emph{Chinese
J. Astron. Astrophys.}, \textbf{2}, 449-453 (2002)

\bibitem{KP03}
P. Kumar and A. Panaitescu, \emph{Mon. Not. R. Astron. Soc.},
\textbf{346}, 905-914 (2003)

\bibitem{MKP04}
E. McMahon, P. Kumar and A. Panaitescu,
\emph{Mon. Not. R. Astron. Soc.}, \textbf{354}, 915-923 (2004) 

\bibitem{PK04}
A. Panaitescu and P. Kumar, \emph{Mon. Not. R. Astron. Soc.},
\textbf{353}, 511-522 (2004)

\bibitem{FZW05a}
Y. Z. Fan, B. Zhang and D. M. Wei, \emph{Astrophys. J.}, \textbf{628},
L25-L28 (2005)

\bibitem{CL00}
R. Chevalier and Z.-Y. Li, \emph{Astrophys. J.}, \textbf{536}, 195-212
(2000) 

\bibitem{wu03}
X. F. Wu, Z. G. Dai, Y. F. Huang and T. Lu,
\emph{Mon. Not. R. Astron. Soc.}, \textbf{342}, 1131-1138 (2003) 

\bibitem{kz03b}
S. Kobayashi and B. Zhang, \emph{Astrophys. J.}, \textbf{597}, 455-458
(2003) 

\bibitem{kmz04}
S. Kobayashi, P. M\'esz\'aros and B. Zhang \emph{Astrophys. J.},
\textbf{601}, L13-L16 (2004)

\bibitem{zwd05}
Y. C. Zou, X. F. Wu. and Z. G. Dai, \emph{Mon. Not. R. Astron. Soc.},
in press, astro-ph/0508602 (2005) 

\bibitem{FZW05b}
Y. Z. Fan, B. Zhang and D. M. Wei, \emph{Astrophys. J.}, \textbf{629},
334-340 (2005)

\bibitem{B05}
A. M. Beloborodov, \emph{Astrophys. J.}, \textbf{618}, L13-L16 (2005)

\bibitem{ZK05}
B. Zhang and S. Kobayashi, \emph{Astrophys. J.}, \textbf{628},
315-334 (2005)

\bibitem{FWW04}
Y. Z. Fan, D. M. Wei and C. F. Wang, \emph{Astron. Astrophys.},
\textbf{424}, 477-484 (2004)

\bibitem{Roming05}
P. Roming et al., \emph{Nature}, submitted, astro-ph/0509273 (2005)

\bibitem{derishev99}
E. V. Derishev, V. V. Kocharovsky and Vl. V. Kocharovsky,
\emph{Astrophys. J}, \textbf{521}, 640-649 (1999)

\bibitem{B03}
A. M. Beloborodov, \emph{Astrophys. J}, \textbf{585}, L19-L22 (2003) 

\bibitem{FZW05c}
Y. Z. Fan, B. Zhang and D. M. Wei, \emph{Astrophys. J.}, \textbf{628},
298-314 (2005)

\bibitem{ZM02}
B. Zhang and P. M\'esz\'aros, \emph{Astrophys. J.}, \textbf{566},
712-722 (2002)

\bibitem{FW04}
Y. Z. Fan and D. M. Wei, \emph{Astrophys. J.}, \textbf{615},
L69-L72 (2004)

\bibitem{rossi04}
E. Rossi, A. M. Beloborodov and M. J. Rees, in \emph{Gamma-Ray Bursts:
30 Years of Discovery}, edited by E. E. Fenimore and M. Galassi, AIP
Conference Proceedings 727, American Institute of Physics, New York,
2004., pp. 198-202 

\bibitem{vestrand05}
Vestrand, W. T. et al., \emph{Nature}, \textbf{435}, 178-180 (2005)

\bibitem{blake05} 
Blake, C. H. et al., \emph{Nature}, \textbf{435}, 181-184 (2005)

\bibitem{blustin05}
A. J. Blustin et al., \emph{Astrophys. J.}, submitted,
astro-ph/0507515 (2005) 

\bibitem{SD05}
L. Shao and Z. G. Dai, \emph{Astrophys. J.}, in press,
astro-ph/0506139 (2005) 

\bibitem{Chincarini05}
G. Chincarini et al.  \emph{Astrophys. J.}, submitted, astro-ph/0506453
(2005) 

\bibitem{Nousek05}
J. A. Nousek et al. \emph{Astrophys. J.}, submitted, astro-ph/0508332
(2005) 

\bibitem{Zhang05}
B. Zhang et al. \emph{Astrophys. J.}, submitted, astro-ph/0508321
(2005) 

\bibitem{Tagliaferri05}
G. Tagliaferri et al. \emph{Nature}, \textbf{436}, 985-988 (2005)

\bibitem{Barthelmy05-1}
S. D. Barthelmy et al., \emph{Astrophys. J.}, submitted (2005) 

\bibitem{KP00}
P. Kumar and A. Panaitescu, \emph{Astrophys. J.},
\textbf{541}. L51-L54 (2000)

\bibitem{FW05}
Y. Z. Fan and D. M. Wei, \emph{Mon. Not. R. Astron. Soc.},
in press, astro-ph/0506155 (2005)

\bibitem{RM94}
M. J. Rees and P. M\'esz\'aros, \emph{Astrophys. J.},
\textbf{430}, L93-L96 (1994)

\bibitem{ZM01}
B. Zhang and P. M\'esz\'aros, \emph{Astrophys. J.},
\textbf{552}, L35-L38 (2001)

\bibitem{RM98}
M. J. Rees and P. M\'esz\'aros, \emph{Astrophys. J.},
\textbf{496}, L1-L4 (1998)

\bibitem{Falcone05}
A. D. Falcone et al., \emph{Astrophys. J.}, submitted (2005) 

\bibitem{Romano05}
P. Romano et al.,  \emph{Astron. Astrophys.}, submitted (2005) 

\bibitem{Kobayashi05}
S. Kobayashi, B. Zhang, P. M\'esz\'aros and D. N. Burrows,
\emph{Astrophys. J.}, submitted, astro-ph/0506157 (2005) 

\bibitem{Gehrels05}
N. Gehrels et al., \emph{Nature}, in press, astro-ph/0505630 (2005)

\bibitem{Barthelmy05-2}
S. D. Barthelmy et al., \emph{Nature}, submitted (2005)

\bibitem{FZP05}
Y. Z. Fan, B. Zhang and D. Proga, 
astro-ph/0509019, (2005)

\bibitem{Proga03}
D. Proga, A. I. MacFadyen, P. J. Armitage and M. C. Begelman,
\emph{Astrophys. J.}, \textbf{599}, L5-L8 (1998)

\bibitem{Wu05}
X. F. Wu, personal communication (2005)

\end{thebibliography}

\IfFileExists{\jobname.bbl}{}
 {\typeout{}
  \typeout{******************************************}
  \typeout{** Please run "bibtex \jobname" to optain}
  \typeout{** the bibliography and then re-run LaTeX}
  \typeout{** twice to fix the references!}
  \typeout{******************************************}
  \typeout{}
 }

\end{document}